\documentclass[preprint]{revtex4-1}
\usepackage{graphicx}
\usepackage{bm}

\begin{document}

\date{\today}

\title{
Probing the magnetoelectric effect in noncentrosymmetric superconductors by equal-spin Andreev tunneling
}  

\author{G. Tkachov}

\affiliation{
Institute of Physics, Augsburg University, 86135 Augsburg, Germany\\
Institute for Theoretical Physics and Astrophysics, University of Wuerzburg, Am Hubland, 97074 Wuerzburg, Germany}

\begin{abstract}
In noncentrosymmetric superconductors (NCSs), the conversion of a charge current into spin magnetization 
 - the so called magnetoelectric effect - is the direct indicator of the unconventional, mixed-parity order parameter. 
This paper proposes a scheme to detect the magnetoelectric effect by anomalous, equal-spin Andreev tunneling in NCS/ferromagnet contacts. 
The proposal relies on the ability to generate spin-polarized triplet pairing by passing an electric current through an NCS. 
Such an induced triplet pairing bears a similarity to the paradigmatic nonunitary pairing in triplet superfluids with a complex vector order parameter ${\bm d}$. 
The qualitative difference is that the induced nonunitary state can be realised in NCSs with a purely real ${\bm d}$ 
by breaking the time-reversal symmetry in current-biased setups. 
This offers a possibility to access the unconventional superconductivity in NCSs through electrical transport measurements. 
\end{abstract}

\maketitle

\section{Introduction}

Superconductivity in materials that lack a center of inversion symmetry does not fit into 
traditional classification of superconducting states invoking definite (even or odd) spatial parity 
(see, e.g., Refs. \cite{Bauer12,Tanaka12,Yip14,Samokhin15,Schnyder15,Smidman17}).
Such noncentrosymmetric superconductors (NCSs) exhibit an antisymmetric spin-orbit coupling (SOC) and, as a result,
the mixing of the even-parity, spin-singlet and odd-parity, spin-triplet Cooper pairs. 
Among intriguing physical consequences of the parity mixing are magnetoelectric effects manifested 
in the conversion of a charge current into spin magnetization 
and vice versa (see, e.g., Refs. \cite{Edelstein95,Yip02,Konschelle15,Tokatly17}), 
the nonuniform (helical) superconducting order \cite{Mineev94,Mineev08} as well as
topological bulk and surface properties (see recent reviews in Refs. \cite{Schnyder15} and \cite{Smidman17}).

\begin{figure}[t]
\begin{center}
\includegraphics[width=75mm]{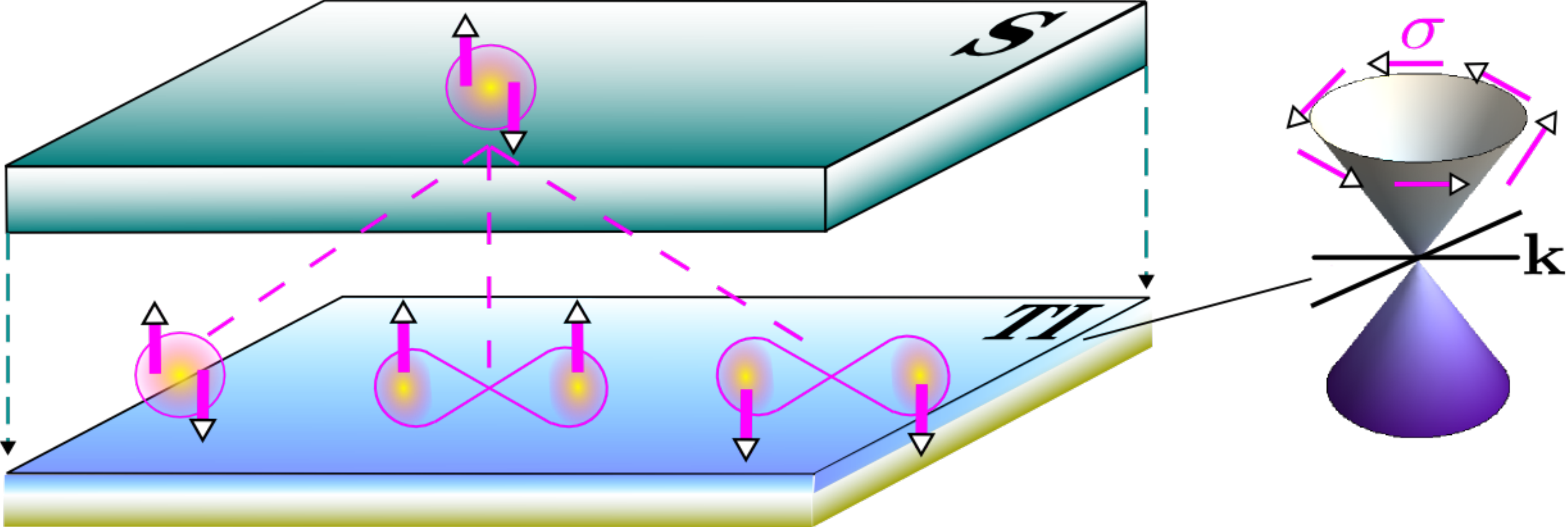}
\vskip 0.5cm
\includegraphics[width=85mm]{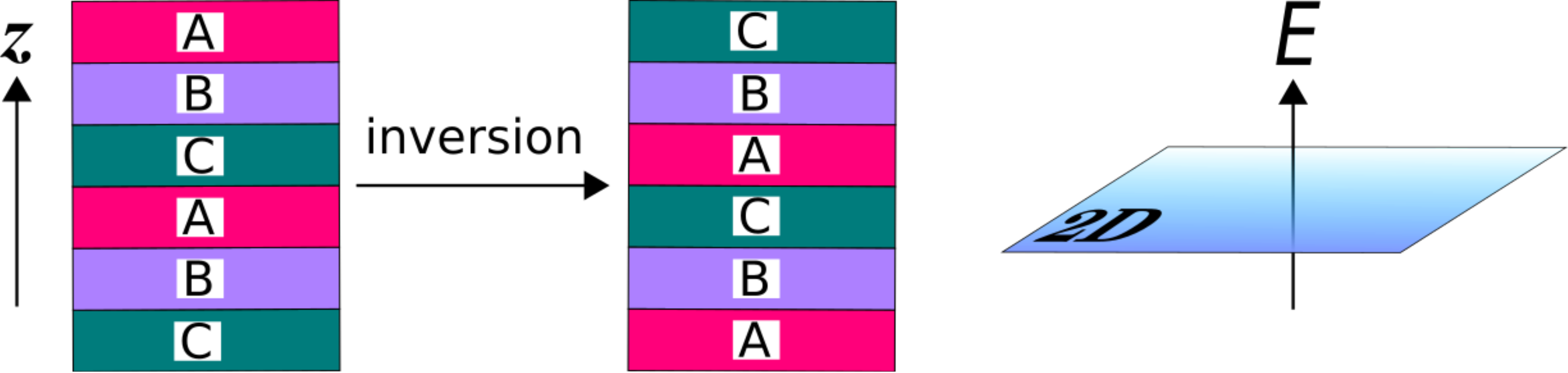}
\end{center}
\caption{
Top: Schematic of a mixed-parity proximity effect in a topological insulator (TI) coupled to a conventional ($s$-wave singlet) superconductor (S). 
The surface spin-momentum-locking facilitates conversion of singlet Cooper pairs into a mixture of singlet and triplet states.
Bottom: Examples of inversion symmetry breaking in NCSs: 
ABC layer stacking (e.g., in tetragonal systems) and two-dimensional systems (interfaces, monolayers and heterostructures) 
in a perpendicular electric field (adapted from Ref. \cite{Sigrist09}).
}
\label{NCSs_fig}
\end{figure}

A close analogue of the intrinsic noncentrosymmetric superconductivity is the mixed-parity superconducting proximity effect 
in hybrid structures of conventional superconductors and normal noncentrosymmetric materials \cite{Edelstein03}. 
It is particularly well pronounced in topological insulators owing to their extraordinary large SOC \cite{Koenig07,Hasan10,Qi11}.
In such proximity structures, the spin-momentum locking converts singlet Cooper pairs into a mixture of induced singlet and triplet states 
\cite{Stanescu10,Potter11,Yokoyama12,BlackSchaffer12,GT13,Snelder15,Reeg15,GT15,Yu16,GT17,Vasenko17a,Vasenko17b,Alidoust17,Hugdal17}.
The induced triplet condensate has no net spin magnetization since, by time-reversal symmetry, 
the pairs with both spin projections $S_z=\pm 1$ occur (see also Fig. \ref{NCSs_fig}). 
Such an induced state resembles intrinsic NCSs lacking a $z \to -z$ inversion symmetry, where the SOC
is described by the Rashba Hamiltonian      
\begin{equation}
H_{\rm so} = {\bm \sigma} \cdot {\bm \Omega}_{\bm k}, \quad {\bm \Omega}_{\bm k} = \alpha_{\rm so} ({\bm k} \times {\bm z}) = \alpha_{\rm so} [k_y, -k_x, 0].
\label{RSOC}
\end{equation}
Above, ${\bm \sigma}$ is the Pauli matrix vector and ${\bm \Omega}_{\bm k}$ is the SOC field depending on the electron wave vector ${\bm k}$;
$\alpha_{\rm so}$ is the SOC constant and ${\bm z}$ is the unit vector in the $z$-direction.

On the experimental side, much effort has been put into proving the parity mixing in intrinsic NCSs, 
using the measurements of the nuclear magnetic resonance relaxation rate, the magnetic penetration depth, thermal conductivity, etc. 
(see recent reviews in Refs. \cite{Yip14} and \cite{Smidman17}).
As for the proximity structures, a number of electron transport measurements have been conducted on topological insulator materials 
(see, e.g., Refs. \cite{D_Zhang11,Koren11,Sacepe11,Veldhorst12,Wang12_STI,Williams12,Maier12,Cho12,Oostinga13,Koren13,Sochnikov15}), 
reporting signatures of the spin-momentum-locked surface states. 
Despite a diverse range of observed properties, the key indicator of the mixed parity in both intrinsic and proximity NCSs 
- the magnetoelectric effect predicted by Edelstein \cite{Edelstein95} - has not been verified yet.
 
This paper is concerned with the observability of the direct magnetoelectric effect in charge transport in NCSs. 
Some earlier theoretical predictions include spin Hall effects and nonequilibrium spin accumulation in superconducting structures \cite{Malshukov08,Malshukov10,Bergeret16,Bobkova16}, 
electrically controllable spin filtering in topological surfaces states \cite{Bobkova17} and magnetoelectric $0 - \pi$ transitions in quantum spin Hall insulators  \cite{GT17_EE}.
The present paper elaborates on an interesting connection between the magnetoelectric effect 
and the nonunitary Cooper pairing in triplet superfluids and superconductors,  
which was pointed out in Ref. \cite{GT17}. 
In nonunitary triplet superfluids and superconductors \cite{Leggett75,Sigrist91,Mackenzie03},
Cooper pairs carry a net spin magnetic moment $\sim i{\bm d}_{\bm k} \times {\bm d}^*_{\bm k}$ 
associated with a complex vector order parameter ${\bm d}_{\bm k}$. 
It turns out that a similar spin-polarized state can be induced in topological insulator/superconductor hybrids by an electric supercurrent \cite{GT17}.
Such a possibility is rather counter-intuitive since in such hybrid structures there is no pairing interaction in the triplet channel, 
hence ${\bm d}_{\bm k}=0$. The role of the ${\bm d}$-vector is assumed by 
the amplitude ${\bm f}(E,{\bm k},{\bm q})$ of the triplet pair correlations induced via the proximity effect in the topological insulator 
(where $E$ is the energy with respect to the Fermi level, while ${\bm q}$ is the phase gradient created by the supercurrent).  
In other words, ${\bm f}(E,{\bm k},{\bm q})$ is a quantitative measure of the singlet-triplet pair conversion  
under the Rashba SOC. Furthermore, the axial vector $i{\bm f}(0,{\bm k},{\bm q}) \times {\bm f}^*(0,{\bm k},{\bm q})$ 
yields the current-induced pair spin polarization at the Fermi level \cite{GT17}:
\begin{equation}
\overline{ 
i 
{\bm f}(0,{\bm k},{\bm q}) \times {\bm f}^*(0,{\bm k},{\bm q})
}\propto \alpha^3_{\rm so} ({\bm q} \times {\bm z}), 
\label{CSP}
\end{equation}
where the bar means averaging over the ${\bm k}$ directions.
This result just means that a charge current is converted into spin magnetization of the superconducting condensate,
a form of the magnetoelectric effect pioneered in normal metals by D'yakonov and Perel' \cite{DP71} and 
studied later in NCSs by Edelstein \cite{Edelstein95} and other authors (see, e.g., Refs. \cite{Yip02,Konschelle15,Tokatly17}). 
The essential difference is that, here, the polarization (\ref{CSP}) is carried by triplet Cooper pairs and 
manifests itself in the Andreev (two-particle) tunneling.

\begin{figure}[t]
\includegraphics[width=70mm]{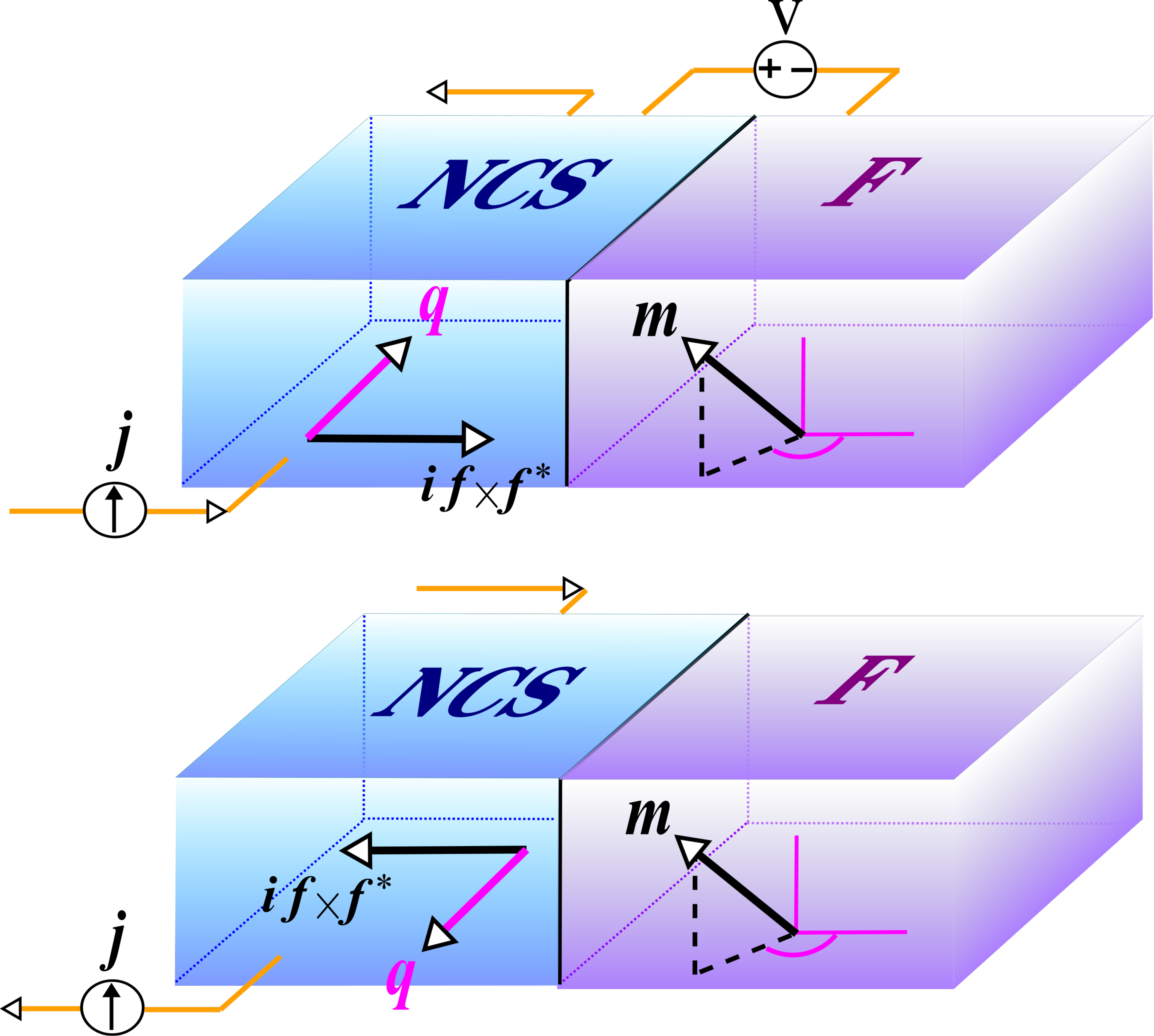}
\caption{
Schematic of the charge-to-spin conversion in an NCS and its detection by a ferromagnetic (F) contact.
An electric current is applied to the NCS, creating a phase gradient ${\bm q}$ parallel to the NCS/F interface.  
The combined effect of the phase gradient and SOC
generates Cooper-pair spin polarization described by the axial vector $i{\bm f} \times {\bm f}^*$.
To detect it, the electric conductance is measured in the direction perpendicular to the interface in the presence of a (small) bias 
voltage $V$. The detection scheme exploits the dependence of the Andreev tunneling on the relative orientation of the pair spin polarization $i{\bm f} \times {\bm f}^*$ 
and magnetization ${\bm m}$ [see also Eq. (\ref{dG_Intro})].
For concreteness, a Rashba NCS is assumed. 
In a cubic NCS, the pair spin polarization would be parallel to ${\bm q}$. 
}
\label{NCS_F_fig}
\end{figure}

The present paper generalizes the above findings beyond the proximity-effect model assumed in Ref. \cite{GT17}. 
Here, we consider intrinsic NCSs, taking into account the triplet pairing interaction represented by a real ${\bm d}$-vector. 
In particular, we focus on two-dimensional (2D) NCSs with the Rashba SOC (\ref{RSOC}) and three-dimensional (3D) NCS 
of the cubic crystal group with the linear SOC: 
\begin{equation}
{\bm \Omega}_{\bm k} = \alpha_{\rm so} {\bm k} = \alpha_{\rm so} [k_x, k_y, k_z].
\label{QSOC}
\end{equation}
To observe the pair spin polarization, we suggest to measure the Andreev conductance of a tunnel junction 
between a current-biased NCS and a ferromagnet. 
The idea is to exploit an analogy with the spin-valve effect, as illustrated in Fig. \ref{NCS_F_fig}. 
Similar to spin valves, the conductance of an NCS/ferromagnet junction should depend on the relative orientation of the magnetization ${\bm m}$ in the ferromagnet
and the average spin polarization $i{\bm f} \times {\bm f}^*$ in the NCS, the latter being controlled by an applied supercurrent. 
The switching of the magnetic configuration of the junction is achieved by reversing the supercurrent direction, 
i.e. by changing the sign of the phase gradient ${\bm q} \to -{\bm q}$.
This produces the change in the tunneling Andreev conductance, 
$G({\bm q}) - G(-{\bm q}) \propto
{\bm m} \cdot 
\overline{
i{\bm f}(0,{\bm k},{\bm q}) \times {\bm f}^*(0,{\bm k},{\bm q})
}
$ \cite{GT17},
which depends strongly on the type of the NCS: 
\begin{eqnarray}
G({\bm q}) - G(-{\bm q}) 
\propto
\left\{
\begin{array}{cc}
{\bm z} \cdot ({\bm m} \times {\bm q}), & Rashba \,\, NCSs,\\ 
({\bm m} \cdot {\bm q}), & cubic \,\, NCSs.               
\end{array}
\right.
\label{dG_Intro}
\end{eqnarray}
The main conclusion is that the Andreev tunneling is asymmetric with respect to the supercurrent direction specified by vector ${\bm q}$.
This asymmetry could serve as a clear smoking gun for the magnetoelectric charge-spin conversion in NCSs, 
which is also interesting in the context of superconducting spintronics \cite{Eschrig11,Linder15,Eschrig15}.
The following sections clarify the details of the theoretical model and methods used and provide extended discussion of the results.   

\section{Induced nonunitary spin-triplet pairing in noncentrosymmetric superconductors}
\label{Sec_NUP}

In this section, we examine the spin-polarized triplet pairing in NCSs carrying a dissipationless electric current. 
There is an instructive parallel here with the paradigmatic nonunitary pairing 
in triplet superfluids and superconductors with a complex ${\bm d}$-vector satisfying $i{\bm d}_{\bm k} \times {\bm d}^*_{\bm k}\not=0$ \cite{Leggett75,Sigrist91,Mackenzie03}. 
To illustrate this point, we adopt the description of the Cooper pairing in terms of 
the Green function of the superconducting condensate
\begin{equation}
\hat{F}(E,{\bm k}) =  [f_0(E,{\bm k}) + {\bm f}(E,{\bm k}) \cdot {\bm \sigma}] i\sigma_y.
\label{F}
\end{equation}
It is written in the standard singlet-triplet spin basis, where  
$f_0(E,{\bm k})$  and ${\bm f}(E,{\bm k})$ are the orbital amplitudes of the singlet and triplet pair correlations, respectively.
The use of the condensate Green function is motivated by two further circumstances: 
$\hat{F}(E,{\bm k})$ is directly related to observable transport properties, and  
it allows for a broader classification of the Cooper pairing which covers 
the hybrid structures \cite{Buzdin05RMP,Bergeret05RMP} and driven superconductors \cite{Triola16}. 

Following Ref. \cite{Leggett75}, we call the pairing nonunitary if the product $\hat{F}\hat{F}^\dagger$ is not proportional to a unit spin matrix. 
Using Eq. (\ref{F}), we find
\begin{equation}
\hat{F}\hat{F}^\dagger = (|f_0|^2 + {\bm f} \cdot {\bm f}^* ) \hat{1} + (f^*_0 {\bm f} + f_0 {\bm f}^*) \cdot {\bm \sigma} + (i{\bm f} \times {\bm f}^*) \cdot {\bm \sigma},
\label{FF}
\end{equation}
where $\hat{1}$ stands for the unit spin matrix (the arguments $E$ and ${\bm k}$ are suppressed for brevity). 
The second and third terms above indicate the nonunitary pairing 
due to the lack of the inversion and time-reversal symmetries, respectively. 
We are interested in the latter case, as it pertains to spin-polarized triplet states.  
We will see that the current-induced pairing and the intrinsic pairing with a complex ${\bm d}$-vector are both characterised 
by a nonvanishing axial vector $i{\bm f} \times {\bm f}^*$ representing Cooper-pair spin polarization. 
The ${\bm f}$-vector formalism allows one to extend the paradigm of the nonunitary pairing beyond its original context, e.g., 
some ferromagnet/superconductor \cite{Fritsch14,Fritsch15} and topological insulator/superconductor \cite{GT17} structures 
can exhibit the spin polarization $i{\bm f} \times {\bm f}^*$, although there is no intrinsic triplet interaction, and ${\bm d}_{\bm k}=0$.  
In such cases, one can speak of an {\em induced} nonunitary pairing. 
The latter can be defined as the triplet pairing with a nonzero spin polarization $i{\bm f} \times {\bm f}^*$ 
induced by a time-reversal symmetry breaking irrespective of the ${\bm d}$-vector. 
The current-induced pairing in NCSs falls precisely into this category because 
it stems from the combined effect of the SOC and supercurrent and does not require a complex ${\bm d}$-vector.
As an example, we consider the NCSs with a real ${\bm d}$-vector proportional to the SO field \cite{Bauer12,Yip14,Samokhin15,Smidman17}:
\begin{equation}
{\bm d}_{\bm k} = \lambda \, {\bm \Omega}_{\bm k},
\label{d}
\end{equation}
where $\lambda$ is the proportionality coefficient. 

\subsection{Condensate Green function for a mixed singlet-triplet pairing interaction}

We begin by briefly reviewing some relevant results for the condensate Green function of a mixed-parity superconductor
in the presence of both SOC and supercurrent \cite{SM_GT17}.  
The normal system is treated as a single-band conductor with the Hamiltonian
\begin{equation} 
\hat{h} = \xi_{\bm k} + {\bm \sigma} \cdot {\bm \Omega}_{\bm k},
\label{h_N}
\end{equation}
consisting of a spin-independent isotropic dispersion $\xi_{\bm k}$ (measured from the Fermi level) and the SOC. 
The superconducting state is described by the Bogoliubov-de Gennes (BdG) Hamiltonian 

\begin{eqnarray}
\hat{H} =
\left[
\begin{array}{cc}
\xi_{{\bm k} + {\bm q}} + {\bm \sigma} \cdot {\bm \Omega}_{ {\bm k} + {\bm q} }  & \hat{\Delta}_{\bm k} \\
\hat{\Delta}^\dagger_{\bm k}  & - (\xi_{-{\bm k} + {\bm q}} + {\bm \sigma} \cdot {\bm \Omega}_{-{\bm k} + {\bm q}})^*
\end{array}
\right],
\,\,\,
\label{H}
\end{eqnarray}
where $\hat{\Delta}_{\bm k}$ comes from the pairing interaction in both even-parity, spin-singlet and odd-parity, spin-triplet channels and is given by
\begin{eqnarray}
\hat{\Delta}_{\bm k} = (\Delta_{\bm k} + {\bm d}_{\bm k} \cdot {\bm \sigma} )i\sigma_y,
\,\,\,
\Delta_{-{\bm k}} = \Delta_{{\bm k}},
\,\,\,
{\bm d}_{-{\bm k}} = -{\bm d}_{\bm k}. 
\label{even_odd}
\end{eqnarray}
The singlet pair potential $\Delta_{{\bm k}}$ is assumed to be isotropic in ${\bm k}$ space. 
Furthermore, the wave-vector shift ${\bm q}$ in Eq. (\ref{H}) accounts for the presence of a superconducting phase gradient 
associated with a dissipationless electric current. It is assumed to be weak enough to disregard the depairing effects in $\hat{\Delta}_{\bm k}$. 
We note in passing that for ${\bm q}=0$ and a ${\bm d}$-vector (\ref{d}) the excitation spectrum consists of two spin-split branches 
\begin{eqnarray}
&
E^{(1)}_{\bm k}=\pm \sqrt{ (\xi_{\bm k} + |{\bm \Omega}_{\bm k}|)^2 + |\Delta_{\bm k} + \lambda  |{\bm \Omega}_{\bm k}| |^2 }, 
&
\nonumber\\
& &
\label{spectrum}\\
&
E^{(2)}_{\bm k}=\pm \sqrt{ (\xi_{\bm k} - |{\bm \Omega}_{\bm k}|)^2 + |\Delta_{\bm k} - \lambda  |{\bm \Omega}_{\bm k}| |^2 }.
&
\nonumber
\end{eqnarray}
The current can be treated as weak if the typical kinetic energy of a Cooper pair, $\hbar v_{_F} |{\bm q}|$, is much smaller than the excitation gaps 
$|\Delta_{\bm k} \pm \lambda  |{\bm \Omega}_{\bm k}||$.    

We are interested in the Green function, $\hat{\cal G}(E,{\bm k})$, of the BdG equation:

\begin{equation}
(E  - \hat{H}) \hat{\cal G}(E,{\bm k}) =  \hat{I},
\label{Eq_G}
\end{equation}
where $\hat{I}$ is a $4\times 4$ unit matrix, and $\hat{\cal G}(E,{\bm k})$ is a matrix of the form

\begin{equation}
\hat{\cal G}(E,{\bm k})=
\left[
\begin{array}{cc}
\hat{G}(E,{\bm k})  &  \hat{F}(E,{\bm k}) \\
\hat{F}^\dagger(E,{\bm k}) &  \hat{\overline{G}}(E,{\bm k})
\end{array}
\right].
\label{G_ph}
\end{equation}
Above, each entry is a $2\times 2$ matrix in spin space: 
$\hat{G}(E,{\bm k})$ and $\hat{\overline{G}}(E,{\bm k})$ are the quasiparticle Green functions, while 
$\hat{F}(E,{\bm k})$ and its hermitian conjugate $\hat{F}^\dagger(E,{\bm k})$ are the anomalous (condensate) Green functions. 
Eq. (\ref{Eq_G}) can be written explictly as  

\begin{eqnarray}
\left[
\begin{array}{cc}
E_+ - {\bm \sigma} \cdot {\bm \Omega}_+  & - (\Delta + {\bm d} \cdot {\bm \sigma} )i\sigma_y \\
i\sigma_y (\Delta^* + {\bm d}^* \cdot {\bm \sigma} ) & E_-  + {\bm \sigma}^* \cdot {\bm \Omega}_-
\end{array}
\right]
\left[
\begin{array}{cc}
\hat{G}  &  \hat{F} \\
\hat{F}^\dagger &  \hat{\overline{G}}
\end{array}
\right]
= 
\hat{I}.
\nonumber
\end{eqnarray}
For ease of calculation we suppress the arguments of the Green functions and introduce the shorthand notations 

\begin{eqnarray}
&
E_+ \equiv E -\xi_{{\bm k} + {\bm q}}, \quad E_- \equiv E + \xi_{-{\bm k} + {\bm q}},
&
\label{E_pm}\\
&
{\bm \Omega}_\pm \equiv {\bm \Omega}_{ \pm {\bm k} + {\bm q} }, \quad \Delta \equiv \Delta_{\bm k}, \quad {\bm d} \equiv {\bm d}_{\bm k}.
&
\label{Om_Delta_d}
\end{eqnarray}
The solution for the condensate function $\hat{F}$ has the general form of Eq. (\ref{F}) (see also Ref. \cite{SM_GT17}),
with the singlet and triplet pair amplitudes defined as

\begin{eqnarray}
&
f_0 =D_0/\Pi, \qquad 
{\bm f} = {\bm D} /\Pi,
& 
\label{f}\\
&
\Pi = (D^2_0 - {\bm D}^2)/(\Delta^2 - {\bm d}^2),
&
\label{Pi}
\end{eqnarray}
where the scalar $D_0$ and vector ${\bm D}$ are given by

\begin{widetext}
\begin{eqnarray}
D_0 &=& \Delta (E_+E_- + {\bm \Omega}_+ \cdot {\bm \Omega}_- - |\Delta|^2) + E_+ {\bm \Omega}_- \cdot {\bm d} + E_- {\bm \Omega}_+ \cdot {\bm d} 
+ i \left({\bm \Omega}_- \times \cdot {\bm \Omega}_+\right) \cdot {\bm d} + \Delta^* {\bm d}^2,
\quad \,\,
\label{D_0}\\
& &
\nonumber\\
{\bm D} &=& 
  (\Delta E_- + {\bm \Omega}_- \cdot {\bm d}){\bm \Omega}_+ 
+ (\Delta E_+ + {\bm \Omega}_+ \cdot {\bm d}){\bm \Omega}_ -  
+ \Delta^2 {\bm d}^*
+ (E_+ E_- - {\bm \Omega}_+ \cdot {\bm \Omega}_- - {\bm d} \cdot {\bm d}^*){\bm d}  
\nonumber\\
& &
\nonumber\\
&+& 
{\bm d} \times ({\bm d} \times  {\bm d}^*)
+
i \Delta\, {\bm \Omega}_+ \times {\bm \Omega}_-
+ iE_+ {\bm d} \times {\bm \Omega}_ -  
- iE_- {\bm d} \times {\bm \Omega}_ +.
\label{D}
\end{eqnarray}
\end{widetext}
Because of the interplay of the ${\bm d}$ vector, SOC and supercurrent, 
the triplet pair correlations have a rather rich structure. 
In particular, the vector cross product terms in Eq. (\ref{D}) are specific to the intrinsic and induced nonunitary pairing.

The case of the intrinsic nonunitary pairing \cite{Leggett75} is recovered by setting the singlet pair potential $\Delta=0$ and SO fields ${\bm \Omega}_\pm =0$ 
in Eqs. (\ref{D_0}) and (\ref{D}) (see also Ref. \cite{SM_GT17}).
Then, for the axial vector $i{\bm f} \times {\bm f}^*$ one readily finds

\begin{equation}
i{\bm f} \times {\bm f}^* = 
\frac{ (E^2 - \xi^2_{\bm k})^2 - \Delta^2_+\Delta^2_-}
{ (E^2 - \xi^2_{\bm k} - \Delta^2_+)^2( E^2 - \xi^2_{\bm k} - \Delta^2_-)^2}
\,\, i{\bm d} \times {\bm d}^*,
\label{ff*_intr_NUP}
\end{equation}
where $\Delta_\pm = \sqrt{ {\bm d} \cdot {\bm d}^* \pm |i{\bm d} \times  {\bm d}^*| }$ 
are the energy gaps for the pairing with the net spin projection parallel ($+$) and antiparallel ($-$) to vector $i{\bm d} \times  {\bm d}^*$.
In this case, the vector $i{\bm f} \times {\bm f}^*$ is simply proportional to $i{\bm d} \times  {\bm d}^*$, carrying essentially the same information. 
An interesting question is whether the polarization $i{\bm f} \times {\bm f}^*$ can occur 
in a superconducting condensate with a purely real ${\bm d}$-vector, where $i{\bm d} \times  {\bm d}^*=0$. 
Such a possibility is rather counter-intuitive, as it does not fit into the standard picture of the nonunitary pairing \cite{Leggett75}. 
Nevertheless, such a possibility does exist in mixed-parity NCSs driven by an electric supercurrent. 
In this case, we encounter an induced triplet state whose formal classification bears a similarity to the intrinsic nonunitary pairing.

\subsection{NCSs with real ${\bm d}$-vectors}

Here, we examine the induced nonunitary pairing in NCSs with a real ${\bm d}$-vector given by Eq. (\ref{d}).
In this case, Eq. (\ref{D}) for vector ${\bm D}$ reduces to

\begin{eqnarray}
{\bm D} &=& 
(\Delta E_-  +  {\bm \Omega}_- \cdot {\bm d}){\bm \Omega}_+  +
(\Delta E_+  +  {\bm \Omega}_+ \cdot {\bm d}){\bm \Omega}_ -  
\nonumber\\
&+&
(\Delta^2  +  E_+ E_-  -  {\bm \Omega}_+ \cdot {\bm \Omega}_-  -  {\bm d}^2){\bm d}
\label{D_real_d}\\
&+&
i \Delta\, {\bm \Omega}_+ \times {\bm \Omega}_-
+ iE_+ {\bm d} \times {\bm \Omega}_ -  
- iE_- {\bm d} \times {\bm \Omega}_ +
.
\nonumber
\end{eqnarray}
To proceed further, we notice that for any linear-in-${\bm k}$ SOC we can write the ${\bm d}$-vector in Eq. (\ref{d}) as
\begin{equation}  
{\bm d} = \frac{\lambda}{2} ({\bm \Omega}_+ - {\bm \Omega}_-),
\qquad
{\bm \Omega}_\pm  = \pm {\bm \Omega}_{\bm k}  + {\bm \Omega}_{\bm q}.
\label{d_pm}
\end{equation}
This helps to cast Eq. (\ref{D_real_d}) into a simpler form 

\begin{eqnarray}
{\bm D} = 
\Delta^\prime  ( E^\prime_- {\bm \Omega}_+ + E^\prime_+ {\bm \Omega}_ -  + i {\bm \Omega}_+ \times {\bm \Omega}_-),
\label{D_real_d_pm}
\end{eqnarray}
with the renormalized scalar factors 

\begin{eqnarray}
&&
\Delta^\prime = \Delta + \frac{\lambda}{2}(E_+ - E_-),
\qquad 
\label{Delta^prime}\\
&&
E^\prime_+ = E_+ - \frac{\lambda}{2\Delta^\prime}(\Delta^2 + E^2_+ - {\bm \Omega}^2_+  - {\bm d}^2), 
\label{E+^prime}\\
&&
E^\prime_- = E_- + \frac{\lambda}{2\Delta^\prime}(\Delta^2 + E^2_- - {\bm \Omega}^2_-  - {\bm d}^2).
\label{E-^prime}
\end{eqnarray}
Now, the ${\bm f}$-vector can be obtained from Eq. (\ref{f}). 
For the purpose of this paper, it suffices to calculate ${\bm f}$ at the Fermi level ($E=0$). 
Restoring the arguments of the functions [see Eqs. (\ref{E_pm}) and (\ref{Om_Delta_d})], we find 

\begin{eqnarray}
&&
{\bm f}(0, {\bm k}, {\bm q}) = \frac{ \Delta^\prime_{{\bm k}, {\bm q}}  }{ \Pi(0, {\bm k}, {\bm q}) }\times
\label{f_real_d}\\
&&
\bigl[
\xi^\prime_{-{\bm k}, {\bm q}}\, {\bm \Omega}_{ {\bm k} + {\bm q} } - \xi^\prime_{{\bm k},{\bm q}} \, {\bm \Omega}_{ -{\bm k} + {\bm q} }
+
i{\bm \Omega}_{ {\bm k} + {\bm q} } \times {\bm \Omega}_{ -{\bm k} + {\bm q} }
\bigr],
\nonumber
\end{eqnarray}
where $\Delta^\prime_{{\bm k}, {\bm q}}$ and $\xi^\prime_{{\bm k},{\bm q}}$ are the renormalized functions

\begin{eqnarray}
\Delta^\prime_{{\bm k}, {\bm q}} &=& \Delta_{\bm k} - \frac{\lambda}{2}(\xi_{{\bm k} + {\bm q}} + \xi_{-{\bm k} + {\bm q}}),
\qquad 
\label{Delta^prime1}\\
\xi^\prime_{{\bm k},{\bm q}} &=& \xi_{{\bm k} + {\bm q}} + \frac{\lambda}{2\Delta^\prime_{{\bm k}, {\bm q}}}
(\Delta^2_{\bm k} + \xi^2_{{\bm k} + {\bm q}} - {\bm \Omega}^2_{{\bm k} + {\bm q}}  - \lambda^2 {\bm \Omega}^2_{\bm k}).
\qquad
\label{xi^prime}
\end{eqnarray}
The scalar factor $\Pi(0,{\bm k},{\bm q})$ is not essential here. 

\begin{figure}[t]
\includegraphics[width=75mm]{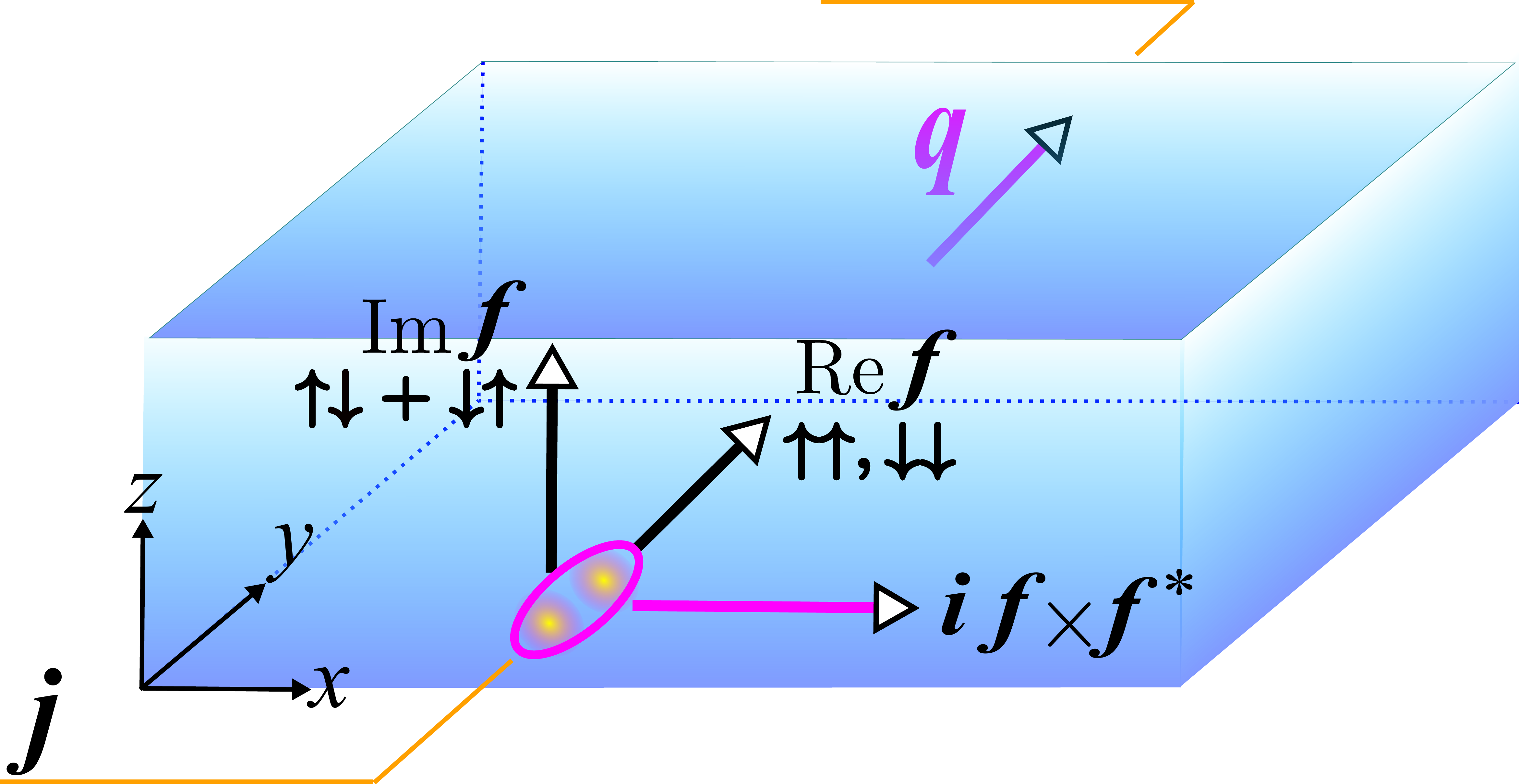}
\caption{
Schematic of the induced nonunitary pairing in a Rashba NCS. The supercurrent generates the imaginary part of the ${\bm f}$ vector 
perpendicular to its real part lying in the SOC plane [see also Eq. (\ref{f_real_d})], 
hence the pair spin polarization $i{\bm f} \times {\bm f}^* = 2 {\rm Re}{\bm f} \times {\rm Im}{\bm f}$. 
In a cubic NCS, the pair spin polarization would be parallel to the phase gradient ${\bm q}$. 
}
\label{NUP_fig}
\end{figure}

Perhaps the most noteworthy feature of the ${\bm f}$-vector (\ref{f_real_d}) is its complex-valuedness. 
The imaginary term in Eq. (\ref{f_real_d}) reflects broken time-reversal symmetry. 
It can be interpreted as the spin {\em torque} exerted by a supercurrent within a Cooper pair.
The torque depends on the misalignment of the SO fields ${\bm \Omega}_{ {\bm k} + {\bm q} }$ and ${\bm \Omega}_{ -{\bm k} + {\bm q} }$ 
acting on the electrons in a Cooper pair.
In the absence of the supercurrent (${\bm q}=0$), the torque vanishes since by time-reversal symmetry the SOC vectors 
${\bm \Omega}_{\bm k}$ and ${\bm \Omega}_{-{\bm k}}$ are antiparallel. 

Furthermore, the imaginary part of the ${\bm f}$-vector (\ref{f_real_d}) is always orthogonal to its real part. 
This is illustrated in Fig. \ref{NUP_fig}. In the case of the Rashba SOC (\ref{RSOC})
both vectors ${\bm \Omega}_{ \pm {\bm k} + {\bm q} }$ lie in the $xy$-plane, so does the real part of the ${\bm f}$-vector. 
It describes a mixture of the equal-spin triplets with the out-of-plane spin projections $S_z = \pm 1$. 
The imaginary part of the ${\bm f}$-vector points out of the SOC plane, yielding the amplitude of the triplet component with $S_z=0$.
In the absence of the supercurrent, the $S_z=0$ triplet is forbidden by time-reversal symmetry.

By analogy with the intrinsic nonunitary states [cf. Eq. (\ref{ff*_intr_NUP})], 
the axial vector $i{\bm f} \times {\bm f}^*$ can be used to characterise the induced pair spin polarization.
From Eq. (\ref{f_real_d}) we find 

\begin{eqnarray}
&&
i{\bm f}(0, {\bm k}, {\bm q}) \times {\bm f}^*(0, {\bm k}, {\bm q}) = 
\frac{
2\Delta^{\prime^2}_{{\bm k}, {\bm q}}
}
{
\Pi(0,{\bm k},{\bm q})^2
}
\times
\label{ff*_real_d}\\
&&
(
\xi^\prime_{-{\bm k}, {\bm q}} \, {\bm \Omega}_{ {\bm k} + {\bm q} } - \xi^\prime_{{\bm k}, {\bm q}} \, {\bm \Omega}_{ -{\bm k} + {\bm q} }
)
\times 
(
{\bm \Omega}_{ {\bm k} + {\bm q} } \times {\bm \Omega}_{ -{\bm k} + {\bm q} }
).
\nonumber
\end{eqnarray}
Clearly, the induced spin polarization is odd in ${\bm q}$. We can therefore linearise it with respect to ${\bm q}$,
using the expansions for ${\bm \Omega}_{ \pm {\bm k} + {\bm q} }$ in Eq. (\ref{d_pm}).  
The result is 

\begin{equation}
i {\bm f}(0,{\bm k},{\bm q}) \times {\bm f}^*(0,{\bm k},{\bm q}) \approx
\frac{8 \Delta^{\prime^2}_{\bm k} \xi^\prime_{\bm k} }{\Pi(0,{\bm k},0)^2}
{\bm \Omega}_{\bm k} \times ({\bm \Omega}_{\bm k} \times {\bm \Omega}_{\bm q}),
\label{ff*_real_d_lin}
\end{equation}
where the renormalized functions $\Delta^{\prime}_{\bm k}$ and $\xi^\prime_{\bm k}$ are given by 
Eqs. (\ref{Delta^prime1}) and (\ref{xi^prime}) with ${\bm q}=0$: 

\begin{equation}
\Delta^\prime_{\bm k} = \Delta_{\bm k} - \lambda \xi_{\bm k}, 
\quad
\xi^\prime_{\bm k} = \xi_{\bm k} + \frac{\lambda}{ 2\Delta^\prime_{\bm k} }
[\Delta^2_{\bm k} + \xi^2_{\bm k}  - (1+\lambda^2) {\bm \Omega}^2_{\bm k}].
\label{Delta_xi^prime}
\end{equation}
Importantly, the spin polarization (\ref{ff*_real_d_lin}) does not vanish upon averaging over the directions of the wave vector ${\bm k}$.
To prove this, we first notice that $\Delta^{\prime}_{\bm k}$ and $\xi^\prime_{\bm k}$ are isotropic 
in ${\bm k}$ space because by assumption $\Delta_{\bm k}$, $\xi_{\bm k}$, and $|{\bm \Omega}_{\bm k}|=\alpha_{\rm so} k$ 
all depend on the wave-vector modulus $k$ only. Moreover, the denominator $\Pi(E,{\bm k},0)$ (\ref{Pi}), which yields the excitation spectrum (\ref{spectrum}), is independent of the ${\bm k}$ direction as well. Then, the angle averaging can be done with the help of the identity 
\begin{equation}
\overline{ \Omega^a_{\bm k} \Omega^b_{\bm k} } = \frac{ 1 }{N}  {\bm \Omega}^2_{\bm k} \delta_{a,b},
\qquad 
a,b=x,y,...,
\label{Om_av}
\end{equation}
which holds for any linear SOC. Above, $\Omega^{a,b}_{\bm k}$ are the Cartesian components of the vector 
${\bm \Omega}_{\bm k}=[\Omega^x_{\bm k}, \Omega^y_{\bm k}, \Omega^z_{\bm k}]$, and
$N$ is the dimensionality of the system, i.e. $N=2$ for Rashba NCSs and $N=3$ for cubic ones.
Besides, for Rashba NCSs the averaging in Eq. (\ref{Om_av}) applies only to the $x$- and $y$-components 
since $\Omega^z_{\bm k}=0$ in this case.

Combining Eqs. (\ref{ff*_real_d_lin}) and (\ref{Om_av}), we arrive at the following result for 
the averaged spin polarization (see also Appendix \ref{A}): 
\begin{eqnarray}
\overline{i {\bm f}(0,{\bm k},{\bm q}) \times {\bm f}^*(0,{\bm k},{\bm q})} =
\frac{ 8(1-N) }{ N }
\frac
{\alpha^2_{\rm so}\Delta^{\prime^2}_k \xi^\prime_k  k^2  }
{\Pi^2_k}
{\bm \Omega}_{\bm q},
\qquad
\label{ff*_real_d_av}
\end{eqnarray}
where we introduce the notations
\begin{equation}
\Delta^\prime_k \equiv \Delta^\prime_{\bm k}, \quad \xi^\prime_k \equiv \xi^\prime_{\bm k}, \quad \Pi_k \equiv \Pi(0,{\bm k},0),
\label{iso}
\end{equation}
emphasizing the isotropic functions in ${\bm k}$ space. 
For the Rashba and cubic NCSs, Eq. (\ref{ff*_real_d_av}) yields

\begin{eqnarray}
&&
\overline{i {\bm f}(0,{\bm k},{\bm q}) \times {\bm f}^*(0,{\bm k},{\bm q})} =
\nonumber\\
&&
=-
\frac
{8 \alpha^3_{\rm so} \Delta^{\prime^2}_k \xi^\prime_k k^2}
{\Pi^2_k}
\left\{
\begin{array}{ccc}
\frac{1}{2} ({\bm q} \times {\bm z}), & Rashba \,\, NCSs, \\
\\
\frac{2}{3} {\bm q},                & cubic \,\, NCSs.
\end{array}
\right.
\label{ff*_real_d_av1}
\end{eqnarray}
Equations (\ref{ff*_real_d_av}) and (\ref{ff*_real_d_av1}) are the main results of this section.  
They show that the induced spin polarization depends sensitively on the type of a structural or lattice asymmetry resulting in the SOC.
Indeed, for Rashba NCSs the spin polarization is induced perpendicularly to an applied supercurrent in the SOC plane [cf. Eq. (\ref{RSOC})], 
while for cubic NCSs it is locked parallel to the current [cf. Eq. (\ref{QSOC})]. 

\section{Equal-spin Andreev tunneling and magnetoelectric effect in NCS/ferromagnet junctions: NEGF approach}
\label{Sec_AR}

This section discusses the observability of the magnetoelectric spin polarization (\ref{ff*_real_d_av}) in Andreev reflection (AR) in NCS/ferromagnet junctions.
It is assumed that the NCS is biased by an electric current applied parallel to the junction interface and
creating the phase gradient, ${\bm q}$, in the direction of the superflow (see also Fig. \ref{NCS_F_fig}). 
The current-induced polarization (\ref{ff*_real_d_av}) is carried by the spin of the AR pairs and, 
because of the spin-charge coupling, produces an odd contribution to the electrical conductance proportional to ${\bm q}$.  
In this respect, the present case differs from the AR spectroscopy of both singlet and triplet superconductors 
(see, e.g., Refs. \cite{Blonder82,Honerkamp98,Hirai01,Asano01,Kastening06,Kowalewski06,Linder07}). 
Also, unlike the above-mentioned theories, to calculate the AR current we use the tunneling Hamiltonian 
in combination with the nonequilibrium Green function (NEGF) method of Ref. \cite{Cuevas96}.
An advantage of this approach is that it does not require a detailed specification of the interface between the normal and superconducting systems, 
allowing for tunnel contacts of different types. 

Summarising the outcome of the NEGF calculation (described in detail elsewhere \cite{Cuevas96,Tkachov02})
we can write the AR current as 
$ 
I_{_A}(V)=(2e/h) \int_{-\infty}^\infty A(E)[n(E-eV)-n(E)]dE, 
$
where $n(E)$ is the Fermi distribution function on the normal side of the junction, $V$ is the voltage difference across the junction, 
and $A(E)$ is the AR probability given by 

\begin{eqnarray}
&
A(E) =
\pi^2 \sum\limits_{{\bm k}_1{\bm k}_2{\bm k}{\bm k}^\prime} 
t_{{\bm k}_1,{\bm k}} t^*_{{\bm k}_2,{\bm k}} \left(t_{{\bm k}_1,{\bm k}^\prime} t^*_{{\bm k}_2,{\bm k}^\prime}\right)^*
&
\nonumber\\
&
\times{\rm Tr}
\left[
\hat{\rho}_{_N}(E,{\bm k}_1) \hat{F}(E,{\bm k}) \hat{\rho}^*_{_N}(-E,-{\bm k}_2) \hat{F}^\dagger(E,{\bm k}^\prime) 
\right].
&
\label{AR_prob}
\end{eqnarray}
Here, $t_{{\bm k},{\bm k^\prime} }$ is the tunneling matrix element, $\hat{\rho}_{_N}(E,{\bm k})$ and $\hat{\rho}^*_{_N}(-E,-{\bm k})$ 
are the spectral functions of a particle and a hole involved in the AR on the normal side [the hole behaves a missing particle propagating back in time], 
and ${\rm Tr}$ means the spin trace operation. Equation (\ref{AR_prob}) holds for a gapped superconductor when the bias and thermal energies are both much smaller than the single-particle excitation gap. For this reason, the AR probability contains only the condensate Green functions of the superconductor,
$\hat{F}(E,{\bm k})$ and $\hat{F}^\dagger(E,{\bm k})$. Also, Eq. (\ref{AR_prob}) admits a straightforward generalization 
in terms of the real-space Green functions. This would not change our results qualitatively, as they reflect the bulk properties of the materials. 

As a suitable observable, we choose the zero-bias and -temperature conductance: $G = \partial I_{_A}/\partial V|_{V,T \to 0} = (2e^2/h) A(0)$. 
It was calculated in Ref. \cite{SM_GT17}, assuming a generic singlet-triplet condensate function [see Eq. (\ref{F})] and 
the Stoner model for the ferromagnet. The calculation is not specific to NCSs, so it suffices to quote the result

\begin{widetext}
\begin{eqnarray}
&&
G = \frac{2e^2\pi^2}{h}\sum\limits_{{\bm k}_1 {\bm k}_2 \alpha=\pm }
\Bigl|
\sum_{ {\bm k} }t_{{\bm k}_1,{\bm k}}t^*_{{\bm k}_2,{\bm k}}
\left[
f_0(0,{\bm k}) + \alpha {\bm f}(0,{\bm k})\cdot {\bm m}
\right]
\Bigr|^2
\rho_\alpha(0,{\bm k}_1)\rho_{-\alpha}(0,-{\bm k}_2)
\label{G_OS}\\
&&
+
\frac{e^2\pi^2}{h}\sum\limits_{{\bm k}_1 {\bm k}_2 \alpha=\pm }
\Bigl\|
\sum_{ {\bm k} }t_{{\bm k}_1,{\bm k}}t^*_{{\bm k}_2,{\bm k}}
\left[
{\bm m} \times ({\bm f}(0,{\bm k}) \times {\bm m})
- i\alpha 
{\bm f}(0,{\bm k}) \times {\bm m}
\right]
\Bigr\|^2
\rho_\alpha(0,{\bm k}_1)\rho_\alpha(0,-{\bm k}_2).
\quad\,\,\,
\label{G_ES}
\end{eqnarray}
\end{widetext}
Above, ${\bm m}$ is a unit vector specifying the magnetization direction in the ferromagnet, 
$\alpha= \pm$ denote the spin projections of the majority ($+$) and minority ($-$) states on the magnetization direction ${\bm m}$, 
and $\rho_\alpha(E,{\bm k})=\delta(E - \eta_{\bm k} + \alpha J)$ is the spectral weight for given spin state $\alpha$ 
[where $\eta_{\bm k}$ and $J$ are the band dispersion and the exchange energy].

Before focusing on the NCSs, let us briefly recapitulate some generic features of the model.  
Eqs. (\ref{G_OS}) and (\ref{G_ES}) describe two AR processes that differ by the relative orientation of the particle and hole spins.  
One is the {\em opposite-spin} AR in which the hole spin projection, $-\alpha$, is antiparallel to that of the particle [see Eq. (\ref{G_OS})]. 
This process creates a Cooper pair with the zero total spin projection on the magnetization direction ${\bm m}$.
If ${\bm f}=0$, this is just the usual AR expressed in the tunneling language.
The other process involves the particle and the hole with equal spin projections, $\alpha$ [see Eq. (\ref{G_ES})]. 
In this case, the Cooper pair is created in a triplet state with the spin projection $\pm 1$ on the magnetization direction ${\bm m}$. 
Such {\em equal-spin} AR takes place for non-collinear vectors ${\bm f}$ and ${\bm m}$  and 
is accompanied by the transfer of the spin angular momentum $\pm \hbar$ and torques on the magnetization.
We note that the tunneling triplet pairs exert torques of the two types, ${\bm m} \times ({\bm f} \times {\bm m})$ and ${\bm f} \times {\bm m}$, 
both being orthogonal to the magnetization.
The equal-spin AR has also been demonstrated in other systems by the real-space BdG method \cite{Wu14,Beiranvand16}  
(see also Refs. \cite{Waintal02,Zhao08,Linder09,Alidoust15,Alidoust18} for superconducting spin transport and torques in other contexts). 

A striking example of AR occurs in junctions between NCSs and half-metallic ferromagnets 
that have only the majority states at the Fermi level.
In this case, the ordinary (opposite-spin) AR is completely suppressed, and the conductance $G$ is entirely due to the equal-spin AR, 
see Eqs. (\ref{G_OS}) and (\ref{G_ES}) with the vanishing minority-carrier spectral weight  
\begin{equation}
\rho_-(E,{\bm k})=0.
\label{G_total}
\end{equation}
The conductance can be written as the sum of even- and odd-parity terms with respect to ${\bm m}$, 
\begin{equation}
G = G_{even} + G_{odd}.
\label{G_total}
\end{equation}
The expressions for $G_{even}$ and $G_{odd}$ appear to be particularly simple for a disordered interface 
with a Gaussian-distributed random matrix $t_{{\bm k},{\bm k^\prime} }$ (see also Ref. \cite{SM_GT17}): 

\begin{eqnarray}
G_{even} &=&
\frac{2e^2}{h} \Gamma^2_+ 
\sum_{\bm k}
\left\|
{\bm f}(0,{\bm k},{\bm q}) \times {\bm m}
\right\|^2,
\label{G_even}\\
G_{odd} &=&
\frac{2e^2}{h}
\Gamma^2_+
\,
{\bm m} \cdot 
\sum_{\bm k}
i{\bm f}(0,{\bm k},{\bm q}) \times {\bm f}^*(0,{\bm k},{\bm q}),
\label{G_odd}
\end{eqnarray} 
where $\Gamma_+$ is the energy scale related to the single-particle tunneling rate for the majority states in the ferromagnet.
Eq. (\ref{G_odd}) proves that the axial vector $i{\bm f} \times {\bm f}^*$ couples directly to the magnetization ${\bm m}$ 
and is, therefore, a valid observable characterizing the magnetism of the Cooper pair condensate.  

The case of NCSs is special because Eqs. (\ref{G_even}) and (\ref{G_odd}) appear to be an even and an odd functions 
of the superconducting phase gradient ${\bm q}$. 
This is easy to see from the expressions for the ${\bm f}$ vector  and pair spin polarization  
obtained earlier [see Eqs. (\ref{f_real_d}) and (\ref{ff*_real_d_av})]. 
The resulting dependence $G({\bm q})$ offers a way to detect the magnetoelectric effect in NCSs by 
measuring the AR conductance versus the bias current.
The detection scheme relies on the fact that $G_{odd}({\bm q})$ makes the total AR conductance $G({\bm q})$ 
asymmetric under a reversal of the current direction, ${\bm q} \to -{\bm q}$.
The difference between $G({\bm q})$ and $G(-{\bm q})$ is simply twice the odd term  
\begin{equation}
\Delta G({\bm q}) =
G({\bm q}) - G(-{\bm q})=2G_{odd}({\bm q}).
\label{dG}
\end{equation}
This result demonstrates the magnetoelectric switching of the AR conductance by the supercurrent \cite{Anisotropy}.
The switching effect can be characterized by the relative change in the conductance

\begin{equation}
\frac{\Delta G({\bm q})}{G(0)} =  \frac{ 2G_{odd}({\bm q}) }{ G_{even}(0) } = 
\frac{ 
2 {\bm m} \cdot 
\sum_{\bm k}
i{\bm f}(0,{\bm k},{\bm q}) \times {\bm f}^*(0,{\bm k},{\bm q}) 
}
{  
\sum_{\bm k}
\left\|
{\bm f}(0,{\bm k},0) \times {\bm m}
\right\|^2
},
\label{dG_rel}
\end{equation} 
where $G(0)=G_{even}(0)$ is the conductance value at zero bias current.
Inserting Eqs. (\ref{f_real_d}) and (\ref{ff*_real_d_av}) into (\ref{dG_rel}) and averaging over the ${\bm k}$ directions (see Appendix \ref{A}), 
we can cast the conductance ratio as 

\begin{equation}
\frac{\Delta G({\bm q})}{G(0)} =  \frac{4 c_{_N} }{  \frac{ 3 - N  }{ 1 - N  }m^2_z - 1} \, ({\bm m} \cdot  {\bm\Omega}_{\bm q}) ,
\label{dG_rel_1}
\end{equation}
where $c_{_N}$ is a constant absorbing the integrals over the wave-vector length $k$:

\begin{equation}
 c_{_N} = 
 \int_0^\infty 
\frac
{\xi^{\prime}_k \Delta^{\prime^2}_k  }
{\Pi^2_k}
\, k^{N+1}  dk
\bigg/
\int_0^\infty 
\frac
{\xi^{\prime^2}_k  \Delta^{\prime^2}_k  }
{\Pi^2_k}
\, k^{N+1}  dk.
\label{varkappa}
\end{equation}
In particular, for the Rashba and cubic NCSs, Eq. (\ref{dG_rel_1}) yields

\begin{eqnarray}
\frac{\Delta G({\bm q})}{G(0)} = - 4\alpha_{\rm so}
\left\{
\begin{array}{ccc}
\frac{  c_{_2} }{ m^2_z + 1 } \, {\bm z} \cdot ({\bm m} \times {\bm q}), & Rashba \,\, NCSs, \\
\\
c_{_3} ({\bm m} \cdot  {\bm q}),         & cubic \,\, NCSs.
\end{array}
\right.
\label{dG_rel_2}
\end{eqnarray}

\section{Discussion and conclusions}
\label{Sec_Conclusions}

The conductance asymmetry in Eq. (\ref{dG_rel_1}) is a signature of the magnetoelectric charge-spin conversion in the NCS
where a dissipationless charge current generates a Zeeman-like field ${\bm\Omega}_{\bm q}$ that couples directly 
to the ferromagnetic magnetization ${\bm m}$. The coupling is mediated by AR involving both charge and spin transfer and
is quite universal in the sense that the material parameters are absorbed into a single constant $c_{_N}$ in the prefactor of Eq. (\ref{dG_rel_1}). 
The constant $c_{_N}$ is of the order of the inverse largest energy of the problem, which is typically the Fermi energy $\mu$.  

A striking feature of the magnetoelectric switching effect is its strong dependence on the relative orientation of the supercurrent and magnetization,
which, in turn, depends on the type of a structural or crystal lattice asymmetry [see Eq. (\ref{dG_rel_2})].  
For the Rashba NCSs, the switching effect is largest when the magnetization is perpendicular to both supercurrent and interface, 
while for cubic NCSs the maximum switching ratio is attained for parallel ${\bm m}$ and ${\bm q}$.
In any case, the relative conductance change can be estimated as
\begin{equation}
\left| \frac{ \Delta G({\bm q}) }{G(0)} \right| \sim  \frac{4 \alpha_{\rm so} |{\bm q}| }{\mu}.
\label{dG_highest}
\end{equation}
In order to have an analytically tractable model, 
we had to work in the linear approximation with respect to the ${\bm q}$ which was treated as a small perturbation. 
However, the fact of the conductance asymmetry is independent of that approximation. 
It is inherent to NCSs where the axial vector $i{\bm f}(0,{\bm k},{\bm q}) \times {\bm f}^*(0,{\bm k},{\bm q})$ must vanish for ${\bm q}=0$ 
by time reversal symmetry. Generally, the conductance ratio (\ref{dG_rel}) can attain sizable values 
provided that no pair breaking takes place and the SOC constant $\alpha_{\rm so}$ is sufficiently large.

In a fully gapped NCS, $|{\bm q}|$ is limited by the value at which the Cooper-pair kinetic energy $\hbar v_{_F} |{\bm q}|$ 
reaches the smallest of the energy gaps, $\Delta_{min}$. Therefore, qualitatively
$\left| \Delta G({\bm q})/G(0) \right| \sim  4 E_{\rm so}  \Delta_{min}/\mu^2$,
where $E_{\rm so} = \alpha_{\rm so} k_{_F}$ is a characteristic band splitting energy.
From the available numbers $E_{\rm so} \approx 100$ meV and $\Delta_{min} \approx 1$ meV 
for fully gapped NCSs \cite{Smidman17}, one arrives at the estimate 
$\left| \Delta G({\bm q})/G(0) \right| \sim  400$ (meV)$^2/\mu^2$.  
That is, the favourable experimental regime is when the Fermi energy $\mu < 100$ meV.

The role of the ferromagnet spin polarization deserves separate comment. 
Although very helpful, the choice of the half-metallic ferromagnets is not crucial for understanding the present work.  
The equal-spin AR and the odd-parity conductance occur for any degree of the spin polarization.
For a partially polarized ferromagnet, there would be an opposite-spin AR contribution to the even-parity conductance, 
which would predictably reduce the conductance ratio (\ref{dG_rel}).
Besides, the odd conductance (\ref{G_odd}) would acquire a contribution from the minority spin subband that 
comes with the negative sign, so the result would be proportional to the ferromagnet spin polarization at the Fermi level.

In conclusion, this paper has established a link between the mixed-parity order parameter, charge-spin conversion and Andreev tunneling in NCSs.    
This could help to identify the symmetry of the order parameter in NCSs, using transport measurements.
The established link between Andreev tunneling and nonunitary triplet pairing is valid beyond the immediate context of this paper.
With some modifications the presented theory can be implemented for transport characterization 
of the unconventional triplet superconducting states in LaNiGa$_2$ (see Refs. \cite{Hillier12,NUP_multiband}) 
and ferromagnetic SrRuO$_3$/Sr$_2$RuO$_4$ junctions (see Ref. \cite{Anwar16}). 

\acknowledgments

The author thanks Professor Yoshiteru Maeno for valuable discussions and the German Research Foundation (DFG) for financial support 
through Grant No TK60/4-1 and TRR80.

\appendix

\section{Calculation of average pair spin polarization (\ref{ff*_real_d_av}) and magnetoelectric switching ratio  (\ref{dG_rel_1}) }
\label{A}

We begin by averaging Eq. (\ref{ff*_real_d_lin}) over the directions of the wave-vector ${\bm k}$:

\begin{equation}
\overline{
i {\bm f}(0,{\bm k},{\bm q}) \times {\bm f}^*(0,{\bm k},{\bm q}) 
}
=
\frac{8 \Delta^{\prime^2}_{\bm k} \xi^\prime_{\bm k} }{\Pi(0,{\bm k},0)^2}
\overline{
{\bm \Omega}_{\bm k} \times ({\bm \Omega}_{\bm k} \times {\bm \Omega}_{\bm q})
}.
\label{ff*_av}
\end{equation}
It is convenient to write the double vector cross product in terms of the scalar ones:

\begin{equation}
{\bm \Omega}_{\bm k} \times ({\bm \Omega}_{\bm k} \times {\bm \Omega}_{\bm q})
=
{\bm \Omega}_{\bm k}  ({\bm \Omega}_{\bm k} \cdot {\bm \Omega}_{\bm q}) - 
{\bm \Omega}^2_{\bm k} {\bm \Omega}_{\bm q},
\label{cross_product}
\end{equation}
and, then, average the Cartesian components of Eq. (\ref{cross_product}) as follows

\begin{eqnarray}
\overline{
{\bm \Omega}_{\bm k} \times ({\bm \Omega}_{\bm k} \times {\bm \Omega}_{\bm q}) \big|^a
}
&=&
\overline{
\Omega^a_{\bm k}  \sum\limits_{b=x,y...} 
\Omega^b_{\bm k} \Omega^b_{\bm q}
}
- 
\overline{
{\bm \Omega}^2_{\bm k} \Omega^a_{\bm q}
}
\label{cross_product_av}\\
&=&
\sum\limits_{b=x,y...}
\overline{
(\Omega^a_{\bm k}   \Omega^b_{\bm k})
}
\,
\Omega^b_{\bm q}
- 
\overline{
{\bm \Omega}^2_{\bm k} 
}
\,
\Omega^a_{\bm q},
\quad
\label{cross_product_av_1}
\end{eqnarray}
where $a$ and $b$ mean $x,y,...$, and $\Omega^{a,b}_{\bm k}$ are the Cartesian components of the SO vector 
${\bm \Omega}_{\bm k}=[\Omega^x_{\bm k}, \Omega^y_{\bm k}, \Omega^z_{\bm k}]$.
Using the identity in Eq. (\ref{Om_av}), we can write the averaged product in Eq. (\ref{cross_product_av_1}) as 

\begin{eqnarray}
\overline{
{\bm \Omega}_{\bm k} \times ({\bm \Omega}_{\bm k} \times {\bm \Omega}_{\bm q}) \big|^a
}
&=&
\frac{1}{N}
{\bm \Omega}^2_{\bm k}
\,
\Omega^a_{\bm q}
- 
{\bm \Omega}^2_{\bm k} 
\,
\Omega^a_{\bm q}
\label{cross_product_av_2}\\
&=&
\frac{1-N}{N}
{\bm \Omega}^2_{\bm k}
\,
\Omega^a_{\bm q}
\label{cross_product_av_3}\\
&=&
\frac{1-N}{N}
\alpha^2_{_{\rm SO}} k^2
\,
\Omega^a_{\bm q}.
\label{cross_product_ab_4}
\end{eqnarray}
Inserting Eq. (\ref{cross_product_ab_4}) back into Eq. (\ref{ff*_av}), 
we find

\begin{equation}
\overline{
i {\bm f}(0,{\bm k},{\bm q}) \times {\bm f}^*(0,{\bm k},{\bm q}) \big|^a
}
=
\frac{8(1-N)}{N}
\frac{\alpha^2_{_{\rm SO}} \Delta^{\prime^2}_{\bm k} \xi^\prime_{\bm k} k^2}{\Pi(0,{\bm k},0)^2}
\Omega^a_{\bm q}.
\label{ff*_av_1}
\end{equation}
Restoring the vector notations and using Eq. (\ref{iso}), we arrive at Eq. (\ref{ff*_real_d_av}) for the average pair polarization discussed in the main text.

Next we turn to the magnetoelectric switching ratio in Eq. (\ref{dG_rel}) of the main text. 
We need to evaluate the ${\bm k}$ integrals in the conductances $G_{odd}({\bm q})$ and $G_{even}(0)$. 
Let us start with the odd-parity conductance

\begin{eqnarray}
&&
G_{odd} ({\bm q})=
\frac{2e^2\Gamma^2_+}{h}
V_{_N}
\int \frac{d^N {\bm k}}{(2\pi)^N}
{\bm m} \cdot 
[i{\bm f}(0,{\bm k},{\bm q}) \times {\bm f}^*(0,{\bm k},{\bm q})]
\nonumber\\
&&=
\frac{2e^2\Gamma^2_+}{h}
V_{_N}
\int\limits_0^\infty \frac{k^{N-1}dk }{ 2\pi^{N-1} }
{\bm m} \cdot 
\overline{
[i{\bm f}(0,{\bm k},{\bm q}) \times {\bm f}^*(0,{\bm k},{\bm q})]
}.
\label{G_odd_k}
\end{eqnarray}
Here, we converted the sum over ${\bm k}$ into an integral in the $N$-dimensional momentum space ($V_{_N}$ is the volume of the real space) and 
separated the integration over the wave-vector length $k$ from the angle averaging (denoted by the bar).
Inserting Eq. (\ref{ff*_real_d_av}) for the average pair polarization into Eq. (\ref{G_odd_k}), we find 

\begin{eqnarray}
G_{odd}({\bm q})=
\frac{8e^2\Gamma^2_+}{h}
\frac{ \alpha^2_{_{\rm SO}} (1-N) V_{_N} ({\bm m} \cdot {\bm \Omega}_{\bm q})}{N}
\int\limits_0^\infty
\frac
{\Delta^{\prime^2}_k \xi^\prime_k  k^{N+1}dk }
{\Pi^2_k \pi^{N-1}}.
\label{G_odd_k_1}
\end{eqnarray}
In the same manner, we can write $G_{even}(0)$:

\begin{eqnarray}
&&
G_{even}(0) =
\frac{2e^2\Gamma^2_+}{h}
V_{_N}
\int \frac{d^N {\bm k}}{(2\pi)^N}
\| {\bm f}(0,{\bm k},0) \times {\bm m} \|^2
\nonumber\\
&&
=
\frac{2e^2\Gamma^2_+}{h}
V_{_N}
\int\limits_0^\infty \frac{k^{N-1}dk }{ 2\pi^{N-1} }
\,
\overline{
\| {\bm f}(0,{\bm k},0) \times {\bm m} \|^2
}.
\label{G_even_k}
\end{eqnarray}
The triplet amplitude ${\bm f}(0,{\bm k},0)$ is given by Eq. (\ref{f_real_d}) for vanishing phase gradient ${\bm q}=0$:

\begin{eqnarray}
{\bm f}(0, {\bm k}, 0) = \frac{2 \Delta^\prime_k  \xi^\prime_k }{ \Pi_k }  {\bm \Omega}_{\bm k}.
\label{f_q_0}
\end{eqnarray}
Hence, 

\begin{eqnarray}
G_{even}(0) =
\frac{4e^2\Gamma^2_+}{h}
V_{_N}
\int\limits_0^\infty \frac{k^{N-1}dk }{ \pi^{N-1} }
\,
\frac{\Delta^{\prime^2}_k  \xi^{\prime^2}_k }{ \Pi^2_k } 
\overline{
({\bm \Omega}_{\bm k} \times {\bm m} )^2
}.
\label{G_even_k_1}
\end{eqnarray}
Then, the angle averaging is done as follows

\begin{eqnarray}
\overline{
({\bm \Omega}_{\bm k} \times {\bm m} )^2
}
&=&  
\overline{ 
{\bm \Omega}^2_{\bm k} 
}
- 
\overline{
({\bm \Omega}_{\bm k} \cdot {\bm m} )^2
}
\label{cross_product_square}\\
&=& 
\overline{ 
{\bm \Omega}^2_{\bm k} 
}
- 
\sum\limits_{a,b=x,y...}
m^a m^b 
\overline{
\Omega^a_{\bm k} \Omega^b_{\bm k}
}
\label{cross_product_square_1}\\
&=& 
{\bm \Omega}^2_{\bm k} 
- 
\frac{{\bm \Omega}^2_{\bm k}}{N}
\sum\limits^N_{a=x,y...}
(m^a)^2 
\label{cross_product_square_2}\\
&=& 
{\bm \Omega}^2_{\bm k} 
\frac{(3-N)m^2_z - (1-N)}{N},
\label{cross_product_square_3}
\end{eqnarray}
where we again invoke the identity in Eq. (\ref{Om_av}).
The case of Rashba NCSs is somewhat subtle, as the system is 2D, and the summation indices in Eq. (\ref{cross_product_square_1}) run over $x$ and $y$ only,
while the magnetization vector ${\bm m}$ is 3D and satisfies $m^2_x + m^2_y + m^2_z=1$.  
Inserting Eq. (\ref{cross_product_square_3}) back into Eq. (\ref{G_even_k_1}), we find

\begin{eqnarray}
G_{even}(0) =
\frac{4e^2\Gamma^2_+}{h}
\frac{ \alpha^2_{_{\rm SO}} [ (3-N)m^2_z - (1-N)] V_{_N}  }{N}
\int\limits_0^\infty
\frac
{\Delta^{\prime^2}_k \xi^{\prime^2}_k  k^{N+1}dk }
{\Pi^2_k \pi^{N-1}}.
\label{G_even_k_2}
\end{eqnarray}
Combining Eqs. (\ref{G_odd_k_1}), (\ref{G_even_k_2}), and (\ref{dG_rel}), we arrive at Eq. (\ref{dG_rel_1}) for the magnetoelectric switching ratio 
discussed in the main text.


\begin{thebibliography}{99}
\bibitem{Anisotropy}
The switching of the ferromagnetic magnetization by the magnetic field of the supercurrent can be prevented by
choosing a material with strong uniaxial magnetic anisotropy.
\end{thebibliography}
\end{document}